# Materials, physics, and systems for multicaloric cooling

*Huilong Hou* [1,5,#,*], *Suxin Qian* [2,#], *Ichiro Takeuchi* [3,4,*]

[1] Institute of Solid Mechanics, School of Aeronautic Science and Engineering, Beihang University, Beijing 100191, People's Republic of China

[2] Department of Refrigeration and Cryogenic Engineering, Xi'an Jiaotong University, Xi'an, Shaanxi 710049, People's Republic of China

[3] Department of Materials Science and Engineering, University of Maryland, College Park, Maryland 20742, United States of America

[4] Maryland Quantum Materials Center, University of Maryland, College Park, Maryland 20742, United States of America

[5] Beihang Hangzhou Innovation Institute (Yuhang District), Hangzhou, Zhejiang 310023, People's Republic of China

[#] Equally contributed.

[*] Correspondence and requests for materials should be addressed to H.H. (huilong_hou@buaa.edu.cn) or I.T. (takeuchi@umd.edu).

(Manuscript was updated on November 22, 2021)

**Abstract |** The calls to minimize greenhouse gas emissions and the demands for higher energy efficiency continue to drive the research worldwide in alternative cooling and refrigeration technologies. The field of caloric cooling has undergone a series of transformations in the past several decades bolstered by the advent of new materials and devices, and these developments have helped fuel the advent of multicalorics in the past decade. Multicaloric materials systems inherently display one or more types of ferroic orders that can give rise to multiple field-induced phase transitions responsible for the caloric effects. Just as multiferroic materials possessing co-existing order parameters have unveiled a plethora of physical phenomena and applications over the years, multicaloric cooling materials have the potential to open up entirely new avenues for extracting heat and spearhead hitherto unknown technological applications. In this Perspective, we survey the emerging field of multicaloric cooling and explore the state-of-the-art caloric materials systems responsive to multiple fields. We present our visions of future applications of multicaloric and caloric cooling in general including key factors governing the overall system efficiency of the cooling devices.

## 1. Introduction

Cooling technologies are ubiquitous in modern society, and they represent a significant fraction of worldwide energy consumption and greenhouse gas emission. Space cooling and refrigeration based on vapor compression consume about 20% of all electricity generated on earth[1]. Leakage of one kilogram of vapor compression refrigerant to the atmosphere is equivalent to two tons of carbon dioxide in global warming potential (GWP), an amount produced by one car running on fossil fuel for six months (See Related Link "Refrigeration and air-conditioning – Consumers"). The two-century-old vapor compression technology is an engineering marvel featuring highly optimized efficiency, but the calls to minimize greenhouse gas emissions and the demands for even higher energy efficiency are driving research in alternative cooling and refrigeration technologies.

Cooling based on caloric materials (magnetocaloric, electrocaloric, and mechanocaloric materials)[2-6] is considered frontrunners in alternative cooling[7,8] owing to their naturally high energy-conversion efficiencies in addition to being entirely green. They derive heat-pumping capabilities from the manipulation of internal order parameters. During the past ten years, the field of caloric cooling has witnessed a steady increase in the variety of available caloric materials and demonstrated prototypes[9-16].

A particularly exciting new direction in the field is multicaloric materials systems[2,3,17-19], which evince ferroic order(s) with the first-order transition in response to more than one external field. Such caloric materials can naturally possess coupled co-existing ferroic order parameters, and an emerging new trend in the field is multicaloric cooling where multiple fields (magnetic, electric, or stress) can be applied in the multicaloric materials to pump heat from low temperature to high temperature[20-26]. In analogy to stress transduction from a magnetostrictive material to a piezoelectric material (or vice versa) in multiferroic composites, there can also be cooling mechanisms and devices based on multicaloric composite materials and structures. We had witnessed a wide range of multiferroic materials with multiple order parameters resulting in various novel physical phenomena and a myriad of applications over the years[27-30]. Similarly, multicaloric cooling materials have the chance to lead to completely new ways of pumping heat. For the rest of the article, we focus on the utility of the caloric effects for cooling.

In this article, we examine the physics of multicaloric cooling with respect to coupled ferroic order parameters in multiferroic materials. We revisit the scope of mechanocalorics with broadened types of deformation modes, which provide diverse options for realizing multicaloric cooling. We show that *all* caloric cooling processes fall into one of four schemes of multicaloric cooling modes depending on whether a material is a single-phase or a composite and whether a single field or multiple fields are required for pumping heat. We discuss how the dissipated percentage of input energy has a direct impact on the fatigue behavior of multicaloric materials. In addressing these issues, we set the stage for the ultimate question: Can multicaloric cooling help improve the performance of caloric cooling technologies?

## 2. Monocalorics as the foundation for multicaloric effects

From the technological standpoint, materials and devices go hand-in-hand along the course of development. In this section, we highlight and capture the important results from both materials and devices. We start with a quick historical overview of the monocaloric materials used for magnetocaloric[9,31], electrocaloric[10,32], and mechanocaloric cooling[11,33,34], which together serve as the basis of multicaloric cooling, and then we review the history of multicalorics. There are excellent earlier reviews on monocaloric effects, which we refer the readers to for details[3,35-39]. Either adiabatic temperature change, $\Delta T_{\text{ad}}$, or isothermal entropy change, $\Delta S$, can serve as the universal metric to quantify the caloric effect in materials. Technically, $\Delta T_{\text{ad}}$ directly correlates materials to systems because $\Delta T_{\text{ad}}$ as the driving force for heat transfer between the solid refrigerant and the working fluid[40,41], is the metric comparing the materials property (experimentally measured $\Delta T_{\text{ad}}$) to the device performance (temperature span). Also, the industry uses $\Delta T_{\text{ad}}$ to guide the system design: if the $\Delta T_{\text{ad}}$ in materials is less than the required temperature span of the system, the methods such as cascading or active regeneration[42] are called in for designs. $\Delta S$ is relevant to the heat pumped in each cycle and bears a material-specific relationship to $\Delta T_{\text{ad}}$. Herein, we use $\Delta T_{\text{ad}}$.

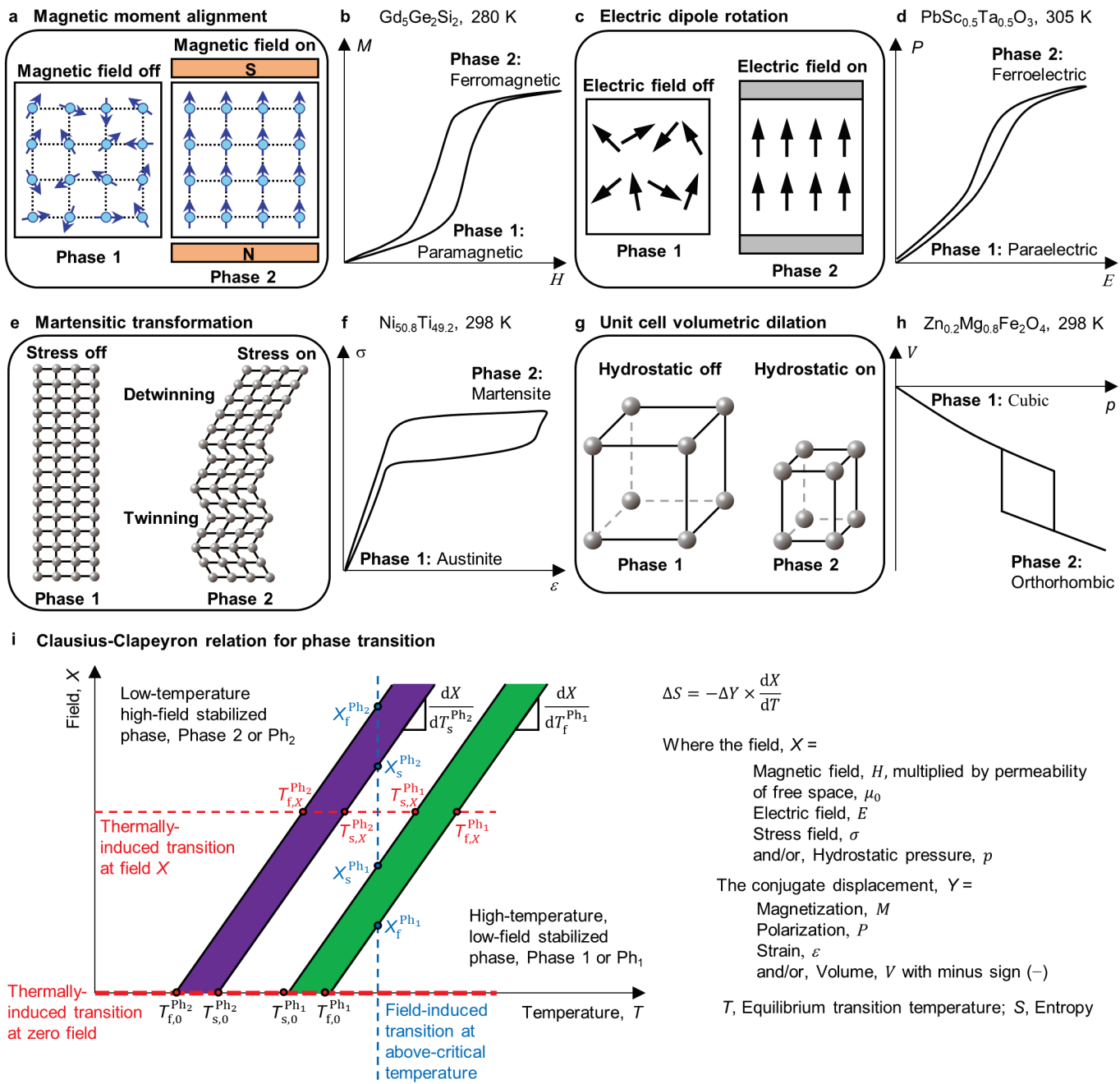

Figure 1 | **Mechanisms of monocaloric materials as the basis for multicaloric cooling.** **a,b** | Schematic of magnetic moments switching between aligned and disordered states upon turning the magnetic field on and off (**a**), manifested by magnetization ($M$) versus applied magnetic field ($H$) at a paramagnetic-ferromagnetic transition (**b**). **c,d** | Schematic of electric dipolar states changing between aligned and randomized upon turning the electric field on and off (**c**), seen in polarization ($P$) versus applied electric field ($E$) at a paraelectric-ferroelectric transition (**d**). **e,f** | Schematic of crystal structure transforming between high-symmetry and twinned-then-detwinned structure as stress is applied and released (**e**), which is described in applied stress ($\sigma$) versus strain ($\varepsilon$) at an austinite-martensite transition (**f**). **g,h** | Schematic of the unit cell volume between dilated and restored—accompanied with lattice parameter change—when hydrostatic pressure is applied and removed (**g**), displayed in volume ($V$) versus applied hydrostatic pressure ($p$) curve at a pressure-induced cubic-orthorhombic transition (**h**). Representative materials at typical test temperatures are shown for each case in **b,d,f,h**. The rotating localized magnetic/electric dipoles drawn in **a** and **c** are but one mode of caloric

cooling effects involving magnetic and electric fields and that we have selected them as representative modes for an illustrative purpose. **i |** Clausius-Clapeyron relation for a phase transition. The high-temperature, low-field stabilized phase, $\text{Ph}_1$, typically possesses a high-symmetry structure such as cubic, and the low-temperature, high-field stabilized phase, $\text{Ph}_2$, has a low-symmetry structure. The $\text{Ph}_1 \leftrightarrow \text{Ph}_2$ transition can be either field-induced at a temperature higher than the transition temperatures, thermally-induced at zero field, or thermally-induced at a finite field with shifted transition temperatures. $X_s^{\text{Ph}_1}$, $X_f^{\text{Ph}_1}$, $X_s^{\text{Ph}_2}$, $X_f^{\text{Ph}_2}$ (where $X$ can be $H$, $E$, $\sigma$, or $p$) are the fields where different phases appear at transition (subscript "s" for start or "f" for finish) for field-induced transition. $T_{s,0}^{\text{Ph}_1}$, $T_{f,0}^{\text{Ph}_1}$, $T_{s,0}^{\text{Ph}_2}$, $T_{f,0}^{\text{Ph}_2}$ are the temperatures where different phases appear under zero field (denoted by subscript "0") whereas $T_{s,X}^{\text{Ph}_1}$, $T_{f,X}^{\text{Ph}_1}$, $T_{s,X}^{\text{Ph}_2}$, $T_{f,X}^{\text{Ph}_2}$ are the temperatures where different phases appear under finite field (denoted by subscript "$X$"). The slope of phase boundaries, $\frac{dX}{dT_f^{\text{Ph}_1}}$, $\frac{dX}{dT_s^{\text{Ph}_2}}$, at above-critical temperatures can be experimentally determined and is equal to $\frac{dX}{dT}$ in Clausius-Clapeyron relation. Panel **b** is adapted with permission from REF.[139], Springer Nature. Panel **d** is adapted with permission from REF.[248], American Physical Society (APS). Panel **f** is adapted with permission from REF.[72], Elsevier. Panel **h** is adapted with permission from REF.[290], APS.

The magnetocaloric effect can be observed in a magnetic material by applying a moment-aligning magnetic field and then removing it, resulting in undoing of the alignment (FIG. 1a), i.e. increase in entropy and absorption of heat from the surrounding during the reversible field-induced process (FIG. 1b). The effect was first discovered in 1917 by Weiss[43] on nickel, and magnetocaloric cooling was proposed in the mid-1920s by Debye[44] and by Giauque[45], and experimentally demonstrated for the first time in 1933 using paramagnetic salts[46]. These early works were carried out at close to zero Kelvin in cryogenic environments. A Nobel Prize in Chemistry was awarded to Giauque in 1949 for his contributions "concerning the behavior of substances at extremely low temperatures"[47]. Active magnetic regeneration near room temperature was first introduced by Brown in 1976 (REF.[48]) where he was able to boost the natural $\Delta T_{\text{ad}}$ of gadolinium (Gd) (14 K, with a field of 7 T) to the regenerator span of 47 K. Giant magnetocaloric materials, Gd–Si–Ge alloys, operating near room temperature was reported in 1997 (REF.[49]). La–Fe–Si alloys[50,51] and magnetocaloric rare-earth-free materials were subsequently discovered including Mn–Fe–P–Si[31] and Ni–Mn-based alloys[9,52]. There have been a number of magnetocaloric cooling prototypes demonstrated over the years including a rotary-type cooler with a temperature span across the regenerator of 10.2 K at a cooling power of 102.8 W and a system COP of 3.1 (REF.[53]).

The electrocaloric effect is achieved by rotating electric dipoles using electric fields (FIG. 1c and d) in much the same way as the magnetic moments are aligned in magnetocaloric materials. An electrocaloric temperature change was reported by Kobeko in 1930 (REF.[54]) and $\Delta T_{\text{ad}}$ of less than 4 mK at 295 K was observed by Wiseman in 1963 (REF.[55]), both on Rochelle salt. In the 1960s, $\Delta T_{\text{ad}}$ of less than 1 K were experimentally measured at cryogenic temperatures on strontium titanate[56,57], cadmium titanate[57], and Li-doped KCl[58,59]. Indirectly measured giant $\Delta S$ due to electric fields were reported in Pb(Zr$_{0.95}$Ti$_{0.05}$)O$_3$ ceramic thin films above room temperature in 2006 (REF.[60]) and poly(vinylidene fluoride-trifluoroethylene-chlorofluoroethylene) [P(VDF-TrFE-CFE) 59.2/33.6/7.2 mole %] polymer thin films near room temperature in 2008 (REF.[32]). $\Delta T_{\text{ad}}$ of 1.2 K in 0.9Pb(Mg$_{1/3}$Nb$_{2/3}$)O$_3$–0.1PbTiO$_3$ (PMN–10PT) bulk relaxor ceramics was boosted by an active regeneration to reach a temperature span of 3.3 K in 2015 (REF.[15]). Multilayer capacitor configurations have been employed to achieve a $\Delta T_{\text{ad}}$ of 3.3 K over a temperature window of 73 K in 2019 (REF.[10]), and to demonstrate regenerative[61] and cascade[62] cooling in 2020. The electrocaloric effect combined with heat transfer enabled by electrostatic actuation has also been reported on polymer films in 2017 (REF.[63]), which was then applied in cascade cooling in 2020 (REF.[64]).

The mechanocaloric effect has primarily been studied in two mechanical-force configurations. One is the elastocaloric effect (FIG. 1e and f), which is most commonly known as the temperature change of an Indian rubber upon stretching and releasing it[65]. In 1859, Joule reported on a range of temperature changes (0.01–1.5 K) exhibited by a series of materials including steel, copper, lead, glass, and wood under uniaxial stress (tension and compression)[66]. The word "elastocaloric" first appeared for Germanium single crystal in 1962 (REF.[67]). In recent years, the elastocaloric effect has primarily been studied in superelastic shape memory alloys including Cu–Al–Ni, Cu–Zn–Al, Ni–Ti, and Ti–Ni–Cu[33,34,68-70]. Some of the largest caloric effects in terms of $\Delta T_{ad}$ and $\Delta S$ have been observed in elastocaloric materials[71-73]. The latent heat of martensitic transformation in shape memory alloys can be as large as 35.1 J g$^{-1}$ (REF.[74]), and at a specific heat capacity of 0.4 J kg$^{-1}$ K$^{-1}$ in $Ni_{49.8}Ti_{30.2}Hf_{20}$, it translates to $\Delta T_{ad}$ as large as 87.8 K, which can be treated as a theoretical upper limit and in practice is higher than the experimentally observed values due to losses. The first elastocaloric cooling system with Ni–Ti wires in a tension mode in a rotating bird-case-like configuration was demonstrated in 2012 (REF.[75]). Recent advances in elastocaloric systems include a compression-based cooling system with heat recovery[76] and a demonstration of a regenerative elastocaloric heat pump[14].

The other popular mechanocaloric cooling mode is the barocaloric effect (FIG. 1g and h), where entropy change is observed in a material upon application of hydrostatic pressure which causes volume change of the materials unit cell often accompanied by lattice distortion. $\Delta T_{ad}$ as large as 9 K has been experimentally measured in poly(methyl methacrylate)[77] with a pressure of 200 MPa at 643 K in 1982. For inorganic rare-earth compounds, $Pr_{0.66}La_{0.34}NiO_3$ with $\Delta T_{ad}$ of 2 K at 300 K under a 1,500 MPa pressure[78] and CeSb with $\Delta T_{ad}$ of 2 K at cryogenic 21 K under a 520 MPa pressure[79] have been experimentally measured in 1998 and 2000, respectively. Indirectly measured $\Delta S$ per mass was recorded near the room temperature in Ni–Mn–In alloys with 27.0 J kg$^{-1}$ K$^{-1}$ under a pressure of 260 MPa in 2010 (REF.[17]), Mn–Ga–N alloys with 22.3 J kg$^{-1}$ K$^{-1}$ under 139 MPa[80] in 2015, and inorganic salts with 60.0 J kg$^{-1}$ K$^{-1}$ under 100 MPa[81] in 2015. The barocaloric effect has also been indirectly measured in perovskite-structure compounds[82,83], superionic conductors[84], plastic crystals (both the reversible endothermic and exothermic processes are shown in REF.[85-87] while only the exothermic process is shown in REF.[88]), and spin crossover compounds[12,89-91].

The parallel development of monocaloric effects has recently converged to the focus point of multicaloric effects. One early development is to observe the different types of monocaloric effects on one same material, where in 1996, elastocaloric $\Delta T_{ad}$ of 5.2 K by varying tensile uniaxial stress of 529 MPa and magnetocaloric $\Delta T_{ad}$ of 8.3 K by varying a magnetic field of 2.5 T has been separately measured in Fe–Rh alloy using a direct method[26]. Later, a vast majority of development in the 2000s makes use of the hydrostatic pressure as a constantly held field to influence the magnetocaloric effect for enhancing the magnitude of entropy change in MnAs compound[92], modifying the order of the transition in $Tb_5Si_2Ge_2$ compound[93], tuning the transition temperatures in $Gd_5Si_2Ge_2$ compound[94], and shifting the operating temperature window in La(Fe, Si)$_{13}$-type compounds[95] and Heusler-type alloys[9]. One theoretical work in 2007 predicts the widening of operating temperature windows by simultaneously varying a magnetic field of 4 T and a pressure of 1.5 GPa in the ErCo$_2$ compound at cryogenic ~30 K[96]. It was not until 2010 that the word "multicaloric" appears for the first time in literature[17]. Since then, the volume of research on multicaloric effect has been substantially increased: theories have been proposed[23,97], type and combination of external fields have been experimentally demonstrated, and the types of the materials responsive to more than one external field have been increasingly explored (See Supplementary Table 1 for a comprehensive list of materials and references).

Over the years' development of multicaloric materials, not only do the types of the materials become diverse but also the configurations have expanded from single-phase materials to composite materials where the constituent components can respond to different types of external fields. Thin films deposited on a substrate is one typical exemplification of the composite materials, where the caloric thin film is subjected to one external field and the substrate to another field as a tuning parameter in a multicaloric cooling cycle as shown in 2016 (REF.[98]), 2019 (REF.[99]), and 2020 (REF.[100]). In the meantime, the composite materials in bulk form of Terfenol-D/Cu–Al–Mn has been studied under one external magnetic field in 2018 (REF.[101]), where one transduction component of Terfenol-D transduces the applied field into the induced field of a different type (mechanical field) and the induced field is used to drive the heat pump cycle (or refrigeration cycle) in the elastocaloric component of Cu–Al–Mn. The strain transduction has also been used in $La_{0.7}Ca_{0.3}MnO_3$ films deposited on $BaTiO_3$ substrates under one applied magnetic field in 2013 (REF.[102]), where the applied field generates a magnetostrictive strain in the film and the strain drives a first-order structural phase transition in the substrate that in turn drives the enhancement of magnetization and entropy changes in the film.

In multicaloric effects, we consider multiple (and sometimes cross-coupled) responses of materials to various fields stemming from the co-existence of magnetic, polar, and structural degrees of freedom associated with phase transitions. The thermodynamics of the caloric effects associated with the first-order phase transitions are well-captured in the Clausius-Clapeyron relation which describes the field-induced $\Delta S$ at a full or partial transition[33,97,103]. Referring to FIG. 1i, the hysteresis of the first-order transition in a material is between a high-temperature, low-field stabilized phase, denoted as phase 1 or $Ph_1$, and a low-temperature, high-field stabilized phase, denoted as phase 2 or $Ph_2$. There are four distinct corresponding temperatures at zero ($T_{s,0}^{Ph_1}$, $T_{f,0}^{Ph_1}$, $T_{s,0}^{Ph_2}$, $T_{f,0}^{Ph_2}$) and at a finite field ($T_{s,X}^{Ph_1}$, $T_{f,X}^{Ph_1}$, $T_{s,X}^{Ph_2}$, $T_{f,X}^{Ph_2}$) and four critical field values at a constant temperature ($X_s^{Ph_1}$, $X_f^{Ph_1}$, $X_s^{Ph_2}$, $X_f^{Ph_2}$), where "s" and "f" in the subscript stand for the start and finish of the phase transition, respectively, and "0" and "X" in the subscript denote at zero and a non-zero field, respectively. At each one of these points, there is a sharp change in a ferroic property during the transition. The field $X$ can be any one of the magnetic field, electric field, stress field, hydrostatic pressure, or their combinations. The slope of phase boundaries, $\frac{dX}{dT}$, is a measure of the sensitivity of critical transformation field on temperature and connects the ferroic properties to the entropy changes in indirect measurements of caloric effects[3]. The phase diagram in a field-temperature space is thus established as a roadmap for the first-order phase transitions in the caloric effect. In the case of multicaloric effects, a combination of fields—for example, $X_1$ and $X_2$ might be magnetic and stress fields, respectively—can be applied simultaneously or sequentially, and the resultant response can be generated "separately" ($\Delta S(X_1) + \Delta S(X_2)$) or "cooperatively" ($\Delta S(X_1, X_2)$).

## 3. Taxonomy of multicaloric cooling

Reflecting the diverse range of functional materials in which multiple ferroic order parameters simultaneously reside, there have been many reports of different types of multicaloric cooling. By now, there have also been numerous reports of composite multicaloric effects where interfacial coupling through transduced stress between materials components is employed as a primary or secondary field in order to pump heat. We take external fields (magnetic, electric, or stress) and constituent materials (upon which the fields are exerted) and classify multicalorics into four major categories (FIG. 2a), where *all* caloric cooling processes can be grouped into based on whether a material is single-phase or composite (FIG. 2b and c) and whether a single or multi-fields are applied for heat pumping (FIG. 2d). In this context, we define multicaloric materials as single-phase materials or composite materials that possess at least one type of ferroic order (lattice, magnetic, and ferroelectric, etc.) and display a caloric effect responsive to more than one type of external field. When the signs of the field-induced caloric effect due to more than one external field are different in a multicaloric material, judicious designs of multicaloric operation are necessary in order to attain the additive instead of subtractive effects. In general, the variation in the magnitude of the caloric effect for a given field is due to the variation of the magnitude of the field in a given direction. The absolute value of the caloric effect increases with the increasing magnitude of the varied field until saturation. Normally, driving by increasing a unipolar field or "undriving" by decreasing the field does not affect this proportionality.

Below we describe each of the four multicaloric categories.

**I. Single field on single-phase materials**

In the multicaloric quad chart (FIG. 2a), the starting single-phase single-field caloric cooling modes are grouped together in Quadrant I, and they are the monocaloric effects, namely, magnetocaloric (MC), electrocaloric (EC), and mechanocaloric (mC, including elastocaloric (eC) and barocaloric (BC)) effects as discussed above (FIG. 2d, I.1–3).

**II. Multiple fields on single-phase materials**

FIG. 2b and c show the multifunctional Venn diagrams of single-phase and composite multicaloric cooling, respectively. Because multiferroic/multicaloric materials can respond to multiple fields, the overlap regions in FIG. 2b represent the first type of multicaloric cooling, single-phase multi-field cooling (Quadrant II in FIG. 2a), the processes of which are schematically shown in FIG. 2d, II.1–4. Region II.1 in FIG. 2b are the single-phase materials where cooling can be induced with either a magnetic field and/or a stress field (FIG. 2d, II.1). For example, $Gd_5(Si_xGe_{1-x})_4$ alloys with $0.24 \leq x \leq 0.5$ exhibit a first-order magnetic transition from paramagnetic to ferromagnetic phases upon cooling while the corresponding structures undergo a transition from a high-temperature monoclinic to a low-temperature orthorhombic phase[104,105], so that the total transition entropy change comprises the entropy change from the structural degree of freedom and the magnetic field-induced entropy change, which add with the same sign. A hydrostatic pressure with a moderate 200 MPa can induce a magneto-structural transition in $Gd_5Si_2Ge_2$ to give rise to an entropy change comparable to the magnetic field (only)-induced entropy change[106]. Another family of materials that belongs to this region is Heusler Ni–Mn–$Z$ ($Z$ = In[107,108], Ga[109], and Sn(Cu)[110]) alloys. Since the structural transition temperature is close to the Curie temperature in these materials and the temperature ranges over which the material exhibits different kinds of ferroic order must also overlap, elastocaloric and magnetocaloric effects simultaneously emerge, where an increase in entropy and a decrease in temperature of the materials can be achieved by applying magnetic field—known as the inverse magnetocaloric effect—or by removing uniaxial stress field—which is the conventional elastocaloric effect. In $Ni_{50}Mn_{35.5}In_{14.5}$, by properly combining a uniaxial stress field of 40 MPa and a magnetic field of 4 T to add the caloric effects cooperatively, its caloric response in $\Delta S$ can be increased from 9.5 to 14 J kg$^{-1}$ K$^{-1}$, an enhancement of 47% compared to its single stimulus-response[107].

An example of a material that belongs to Region II.2 in FIG. 2b is a correlated oxide $VO_2$, where cooling can be induced by electric and/or uniaxial stress fields (FIG. 2d, II.2)[111,112]. The famous metal-insulator transition in $VO_2$ with the crystal symmetry changing from a tetragonal (rutile) structure in a metallic state to a monoclinic structure is accompanied by a latent heat as large as 51.5 J g$^{-1}$ (REF.[113]). Using a relatively low electric field of $7.5 \times 10^{-3}$ MV m$^{-1}$, $\Delta S$ of 94 J kg$^{-1}$ K$^{-1}$ has been achieved in a 0.4-mm-thick $VO_2$ pellet[111], while exerting a uniaxial stress field of 300 MPa in 6-mm-diameter $VO_2$ powders has resulted in a $\Delta T_{ad}$ of 1.6 K[112]. An example of material in Region II.3 of FIG. 2b (and 2d) are $Pb(Fe_{0.5}Nb_{0.5})O_3$–$BiFeO_3$-based ceramics[114], where an electric field of 15 MV m$^{-1}$ leads to a directly measured $\Delta T_{ad}$ of 1.5 K at room temperature and 1.8 K at 348 K while a magnetic field of 9 T has resulted in an indirectly measured $\Delta T_{ad}$ of 0.3 K at a cryogenic temperature of 3 K. To the best of our knowledge, there has not been an experimentally reported demonstration of a multicaloric material response in Region II.4 of FIG. 2b (and 2d), but simulations have shown that simultaneously applying electric, magnetic, and stress fields on $BiFeO_3$ can give rise to a $\Delta T_{ad}$ of 12 K (REF.[115]), while in epitaxial $EuTiO_3$ thin films[116], applying an electric field of 10 MV m$^{-1}$ at a constant 200 MPa is predicted to result in $\Delta T_{ad}$ of 5.31 K and applying a magnetic field of 5 T alone is predicted to lead to a $\Delta T_{ad}$ of 17.3 K. Additional examples with the directionality of the applied fields and the signs of the field-induced caloric effects can be found in the Category II of Supplementary Table 1.

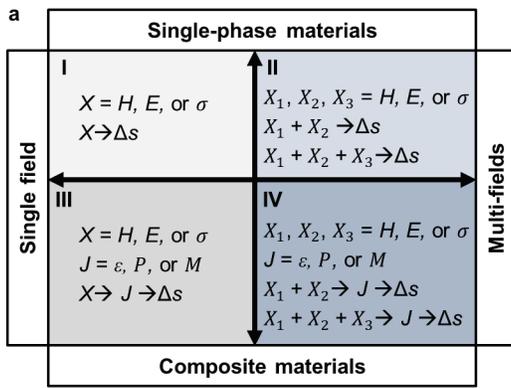
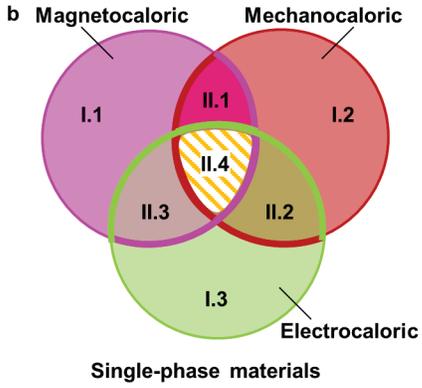
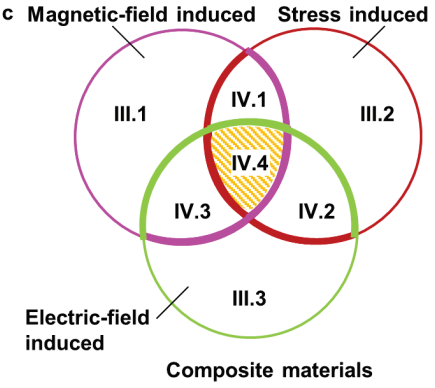
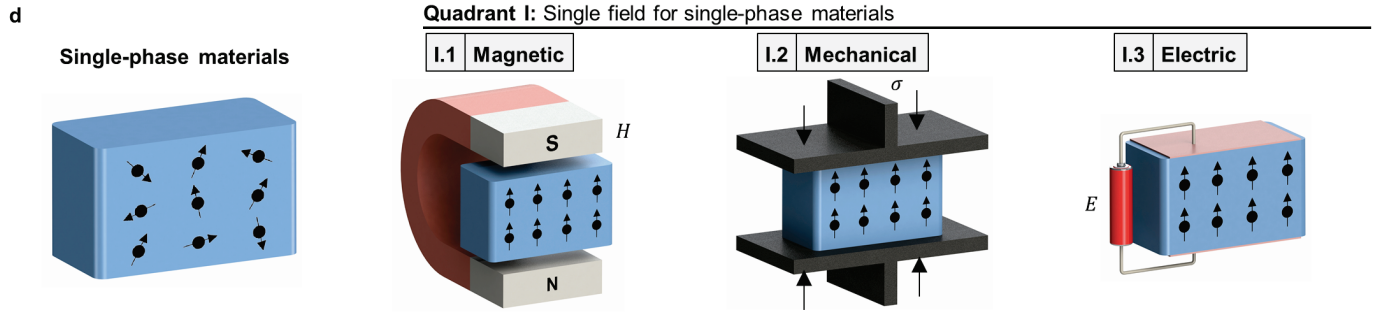
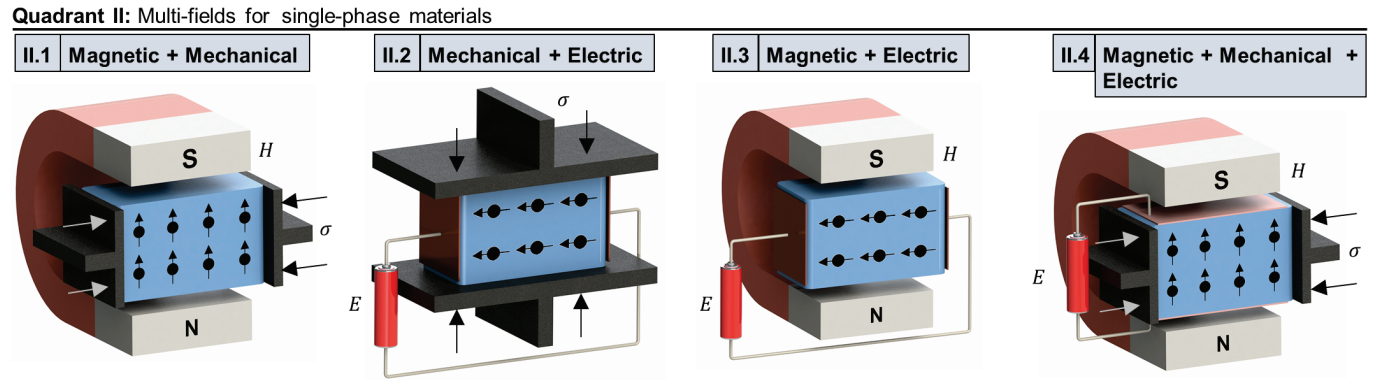
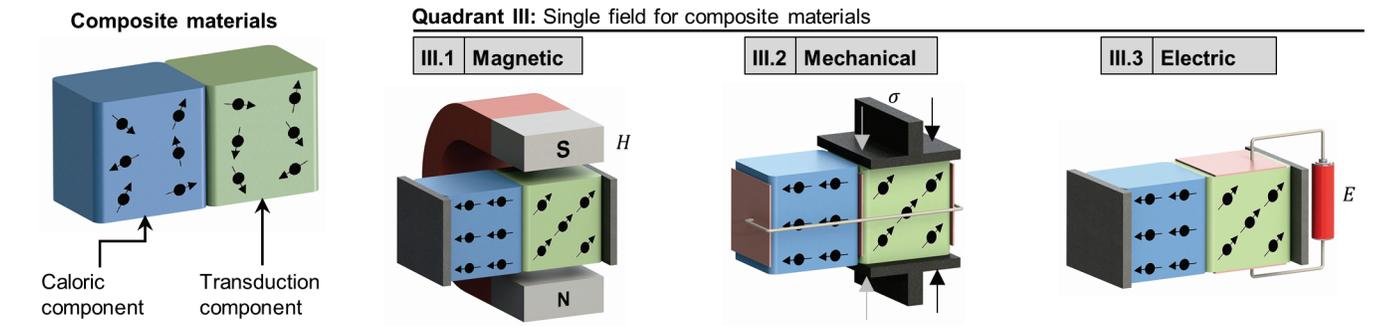
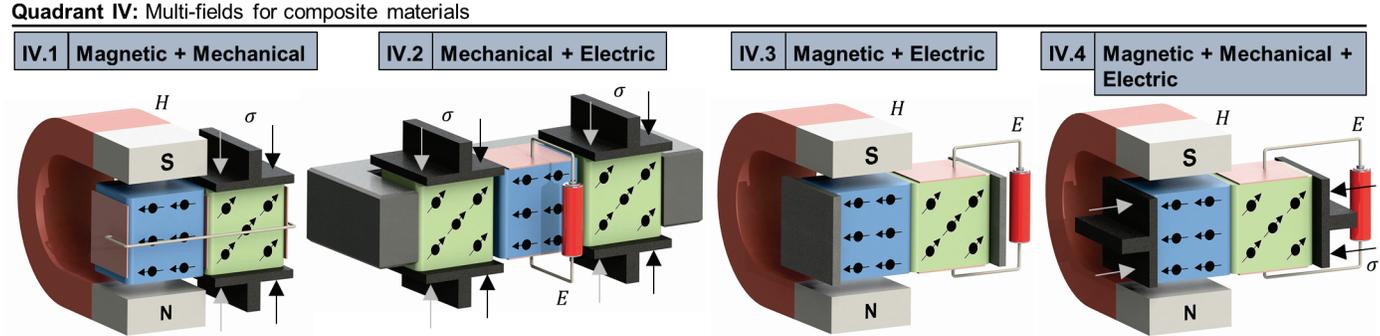

Figure 2 | **Categories of multicaloric cooling**. **a |** Multicaloric quad chart. All monocaloric and multicaloric materials and configurations can be categorized into four quadrants based on whether the materials are single-phase or composite materials and whether single or multiple fields are applied. In Quadrant I, one field is applied to a single-phase material. This is the starting caloric process. In Quadrant II, multiple fields are applied to a single-phase material. In Quadrant III, one field is applied to a composite material. In Quadrant IV, multiple fields are applied to a composite material. The symbol $X$ with a subscript 1, 2, 3 represents different fields (magnetic ($H$), electric ($E$), or stress ($\sigma$)). Composite materials typically involve an indirect transfer, $J$, which is mostly a strain ($\varepsilon$) and in certain cases can be magnetization ($M$) or polarization ($P$). $\Delta s$ denotes the specific entropy change. In multi-field configurations, fields can be applied simultaneously (e.g. $X_1$ and $X_2$), or sequentially (e.g. $X_1$ first, then $X_2$). **b |** Venn diagrams illustrating the relation between magnetocaloric, mechanocaloric, electrocaloric, and multicaloric cooling processes in single-phase materials. **c |** Venn diagram illustrating the relation between magnetic-field induced, stress-induced, and electric-field induced processes in composite materials. **d |** Schematic configurations of single-phase materials and basic caloric processes, I.1 to I.3; single-phase materials under multiple fields, II.1 to II.4; composite materials under single field, III.1 to III.3; and composite materials under multiple fields, IV.1 to IV.4. The ferroic order parameters within materials are denoted by arrows with black circles drawn on materials and are meant to denote the state of any one of the ferroic order parameters ($\varepsilon$, $M$, or $P$). See Supplementary Figure 2 and Supplementary Note 1 for additional configurations of composite materials. For Quadrant IV, four configurations are drawn instead of drawing all possibilities. The method for varying the magnetic field can be by changing the electric current for electromagnets and superconducting magnets at physical positions, or by moving permanent magnets with respect to the materials. The directionality of the different fields in each case represents one of the embodiments.

### III. Single field on composite materials

The next type of multicaloric cooling is composite configurations as shown in the three major circles in FIG. 2c, where one component in the material provides the means to deliver the field needed to pump heat in the other: this is the case of a single field on a composite material which is Quadrant III in FIG. 2a and the circles (and their overlap regions) correspond to different specific modes of cooling depicted in FIG. 2d, III and IV. Given the diversity of transduction (and multiferroic) materials, where the application of one external field leads to the appearance of cross-coupled response, the heat-pumping transduced field can be a magnetic field (due to magnetization), stress field (due to strain at interfaces), or electric field (due to accumulated charges or polarization), as captured in FIG. 2d, III.1–3, respectively. We use $J$ (in FIG. 2a) to denote any of the conjugate displacements (strain $\varepsilon$, polarization $P$, or magnetization $M$) due to corresponding fields. However, we expect the most prevalent transduction mode to be stress (due to $J = \varepsilon$). One demonstrated example is magneto-elastocaloric cooing (FIG. 2d, III.1) where a magnetostrictive Terfenol-D is placed in contact with an elastocaloric Cu–Al–Mn alloy[101]. Here, an ultra-low magnetic field of 0.16 T was able to induce $\Delta T_{ad}$ as high as 4 K, which represents a multi-fold enhancement in magnetic field cooling strength (defined as $\Delta T_{ad}$/applied field) compared to that of "pure" magnetocaloric cooling of FIG. 2d, I.1. This composite effect is able to overcome the large field requirement of the usual single-phase single-field magnetocaloric cooling, as well as the large stress requirement of the usual elastocaloric cooling in FIG. 2d, I.2, but the fact that only one component in the composite undergoes cooling necessitates unique application implementation of this effect such as remote localized cooling of electronic devices[117]. A straightforward extension of this design would be an electro-elastocaloric effect (FIG. 2d, III.3), where piezo-induced strain would be used for the stress application on the elastocaloric material as discussed in REF.[101]. As a variation of this approach, one can consider a mechano-magnetocaloric composite effect (FIG. 2d, III.2), where an application of stress to a magnetostrictive material leads to magnetization alignment, which in turn is used as a source of a magnetic field for a magnetocaloric material placed in contact. Of note is that in composite multicalorics, if the caloric component has heat exchange with the transduction component, heat-insulating objects that are mechanically stiff to deliver displacements should be implemented in between such as a ceramic disk[101].

**IV. Multiple fields on composite materials**

With the three major circles in FIG. 2c corresponding to three modes of single-field composite cooling, the overlapping areas then represent multi-field embodiments of composite multicaloric cooling (Quadrant IV in FIG. 2a). Despite the added complexity, given a diverse choice of materials and fields, it is natural to consider the effect of combinations of multiple fields on composite materials (FIG. 2d., IV.1–4). There have already been several demonstrations of magnetocaloric Fe–Rh thin films coupled elastically to piezoelectric substrates (FIG. 2d, IV.3). In one example, a $Fe_{50}Rh_{50}/BaTiO_3$ composite has seen a nominal 96% reduction of hysteretic losses in the magnetocaloric Fe–Rh film via strain-mediation from the substrate in a multi-step cycle[98], although the dissipated energy associated with the hysteresis still exists and has been transferred from magnetic cycle to elastic cycle. In $Fe_{50}Rh_{50}$/PMN–30PT composites where an electric field is applied to (001)- or (011)-oriented single crystals of PMN–30PT, the Fe–Rh film was shown to undergo a phase transition influenced by the domain switching in the PMN–30PT substrate underneath, and as a result, the magnetic field-induced $\Delta S$ and refrigeration capacity[100], and the operating temperature window of the magnetocaloric effect in the Fe–Rh film[99] has seen significant improvement in addition to the aforementioned hysteresis reduction. By switching on an electric field of $125 \times 10^{-3}$ MV m$^{-1}$ (REF.[118]), a bilayer composite of 0.2-mm-thick $Fe_{48}Rh_{52}$ and 0.2-mm-thick $Pb(Zr_{0.53}Ti_{0.47})O_3$ in a varying magnetic field of 0.62 T has displayed an increase in its transformation temperatures by 2.7 K and reduced nominal thermal hysteresis by 33% from 4.5 K to 3 K despite the existing dissipated energy associated with the hysteresis. In another example, a composite of 12 wt.% magnetocaloric/magnetostrictive $Gd_5Si_{2.4}Ge_{1.6}$ microparticles embedded in a piezo- and pyroelectric poly(vinylidene) fluoride (PVDF) matrix has exhibited a decreased $\Delta S$ of 2.3 J kg$^{-1}$ K$^{-1}$ under application of a magnetic field of 5 T at 265 K compared to the $\Delta S$ of 2.6 J kg$^{-1}$ K$^{-1}$ in pure $Gd_5Si_{2.4}Ge_{1.6}$ microparticles[119]. The difference is due to the interaction of $Gd_5Si_{2.4}Ge_{1.6}$ with the PVDF matrix, indicating that the use of an additional electric field on the composite can potentially provide another knob to tune the multicaloric effect. To the best of our knowledge, there have not been reports of demonstrations of multicaloric configurations in Regions IV.1, 2, and 4 of FIG. 2b (and 2d) either theoretically or experimentally, but it is expected that such embodiments will be reported in the near future. As a straightforward embodiment, in the above-mentioned composite magneto-elastocaloric configuration, one can opt to apply both a magnetic field (to the Terfenol-D piece) and a stress (to the elastocaloric component) in order to enhance the effect.

Although the choice of the pairing of piezoelectricity, piezomagnetism, electrostriction, magnetostriction, or others[27] in the combination of multiple conjugating external fields provides a rich playground, construction of caloric systems that require multiple fields might be limited by their thermodynamic performance and the complexity and cost of the system integration. However, they may provide niche solutions for specialized applications, for example, when one is able to reduce the necessary magnitude of one expensive field by applying another field, where, for instance, a caloric effect driven by the magnetic field from an expensive permanent magnet configuration with no mechanical field can be equivalently driven by a combination of the magnetic field from an inexpensive permanent magnet configuration and a mechanical field from a relatively inexpensive off-the-shelf mechanical actuator[120], or when it is necessary to expand the operating temperature window by applying multiple fields.

**4. Various flavors of mechanocalorics as building blocks for multicalorics**

As we saw above, stress transduction and/or mechanocaloric effects often play an important role in various multicaloric cooling configurations. We, therefore, review different mechanocaloric modes in detail here. The mechanocaloric effect has been primarily demonstrated in magnetic and non-magnetic shape memory alloys[37,121,122] although there have also been reports on other types of materials including polymers[123-125] and ceramics[126,127]. Non-magnetic alloys such as Ni–Ti show ductility, while magnetic alloys such as Ni–Mn–Ga typically exhibit brittleness. The underlying mechanism of the shape memory effect is a diffusionless martensitic phase transformation, and the shape deformation accompanying a martensitic transformation in an unconstrained single crystal has the well-known characteristics of the invariant–plane strain, by which there exists an unrotated and undistorted plane adjoining the transformed and untransformed phases[128]. The invariant–plane strains are manifested in three forms: dilatational strain (responsible for volume change of crystal), shear strain (responsible for shape change of crystal), and their combination. When a lattice transformation from a parent crystal to a product crystal requires an invariant–line (unrotated and undistorted) strain such as one from body-centered cubic (BCC) to face-centered cubic (FCC)[129], the invariant–plane strain at macroscopic shape deformation needs to be intersected by a second invariant–plane strain that can be realized by lattice invariant shear through slip or twinning[130].

From a mechanics point of view, a dilation of an infinitesimal volume element at a point in a solid material stems from isotropic stress, namely, hydrostatic pressure, while a shear strain can be generated by resolving applied stress into shear stress along a projected direction. Dilatational strain and the insensitivity of brittleness to isotropic stress has endowed hydrostatic pressure as a means for a caloric effect with the name "barocaloric", while shear strain from uniaxial stress can lead to entropy change during transformation resulting in an "elastocaloric" effect. To date, most applied stress in reported elastocaloric effects are uniaxial, either tensile or compressive, probably due to the fact that uniaxial stress is the simplest example of deviatoric stress which can be applied in laboratory testing.

Deviatoric stress, however, represents a departure in a stress state from hydrostatic pressure, and in addition to uniaxial stress, there can be a wide variety of stress states. We enlist different elemental forms of deviatoric stresses and define the corresponding mechanocaloric effect for each (panel **a–f** of the figure within BOX 1). We use the term "deviatocaloric" to collectively describe the caloric effects associated with deviatoric stresses. The uniaxial-stress-based elastocaloric effect, which first appeared in literature in 1962[67] as reviewed in Section 2, is but one type of deviatocaloric effect. In fact, *any* stress state in a solid body can be decomposed into hydrostatic and deviatoric stress components. Thus, barocaloric and deviatocaloric effects together make up the family of mechanocaloric effects (Box 1). Recently, there have also been reports of large caloric effects using different deviatoric stresses including torsional stress[131,132] and bending stress[133-136]. In the case of torsional stress, the effect as large as 20.8 K in $\Delta T_{\text{ad}}$ have been observed[131], while bending stress has displayed a $\Delta T_{\text{ad}}$ as large as 15.4 K. Given that these reports with large caloric effects have only appeared in the last three years, it is likely that the other deviatoric effects are being explored or will be reported soon. Demonstration of hitherto unexplored deviatoric caloric modes might also lead to the development of new materials.

Box 1 | **A family of mechanocaloric processes arising from different stress and deformation modes in materials**

Many configurations of the composite multicalorics in FIG. 2 use mechanical transduction, and herein we fully explore all possible mechanocalorics. The most common mechanocaloric effects are elastocaloric (by uniaxial stress) and barocaloric (by isotropic stress) effects. Additional family of processes join the elastocaloric effect and collectively constitute the "deviatocaloric" effects which reflect various forms of deformation and stress distribution. Four other elemental types of deviatocaloric effects are twistocaloric (by torsional stress), flexocaloric (by bending stress), shearocaloric (by shear stress), and hollocaloric (by plane stress). The various family processes are displayed in the figure within Box 1.

In general, the stress state of a point in a solid body is expressed in a tensor notation as

$$\sigma_{ij} = \begin{bmatrix} \sigma_{xx} & \sigma_{xy} & \sigma_{xz} \\ \sigma_{xy} & \sigma_{yy} & \sigma_{yz} \\ \sigma_{xz} & \sigma_{yz} & \sigma_{zz} \end{bmatrix}$$

where the subscripts denote normal stress if they are the same, otherwise they denote shear stress in a rectangular coordinate system *xyz*. The stress state can be decomposed as $\sigma_{ij} = \check{\sigma}_s + \tilde{\sigma}_d$, where $\check{\sigma}_s$ is a spherical stress tensor,

$$\check{\sigma}_s = \begin{bmatrix} \sigma_m & 0 & 0 \\ 0 & \sigma_m & 0 \\ 0 & 0 & \sigma_m \end{bmatrix} \qquad \sigma_m = \frac{1}{3}(\sigma_{xx} + \sigma_{yy} + \sigma_{zz})$$

and $\tilde{\sigma}_d$ is a deviatoric stress tensor.

$$\tilde{\sigma}_d = \begin{bmatrix} \sigma_{xx} - \sigma_m & \sigma_{xy} & \sigma_{xz} \\ \sigma_{xy} & \sigma_{yy} - \sigma_m & \sigma_{yz} \\ \sigma_{xz} & \sigma_{yz} & \sigma_{zz} - \sigma_m \end{bmatrix}$$

The spherical stress tensor, $\check{\sigma}_s$, is a hydrostatic state and responsible for isotropic, volumetric deformation. The deviatoric stress tensor, $\tilde{\sigma}_d$, measures a deviation from the hydrostatic state and causes shape changes. The barocaloric process corresponds exclusively to $\check{\sigma}_s$, while the elastocaloric process is only one scenario of $\tilde{\sigma}_d$; $\tilde{\sigma}_d$ produces other elemental processes and many combined processes. The deformation mode, the resulting stress distribution on the cross section and infinitesimal cube-shaped volume element, and the corresponding stress tensor of barocalorics and five elemental deviatocalorics are shown in the figure within Box 1. Other symbols in the figure are: normal force, $F_n$; torsional moment, $M_t$; bending moment, $M_b$; shear force, $F_s$; gauge pressure, $p$; normal stress, $\sigma$; and shear stress, $\tau$. The formula to compute stress from internal loading can be found in Supplementary Table 2.

Caloric effects through complex deformation modes can be achieved by combining the elemental types. For example, elasto-twistocalorics can be realized by combining the elastocalorics and twistocalorics via applying a set of uniaxial force and torsional moment.

Box 1 | **(Continued)**

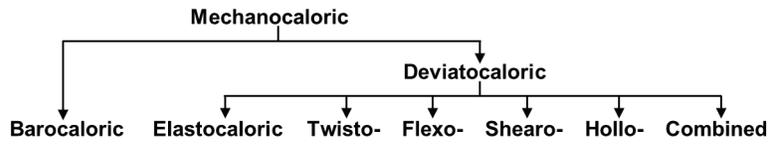

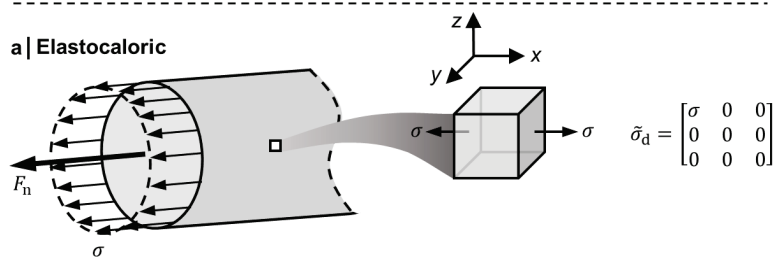

**a | Elastocaloric**

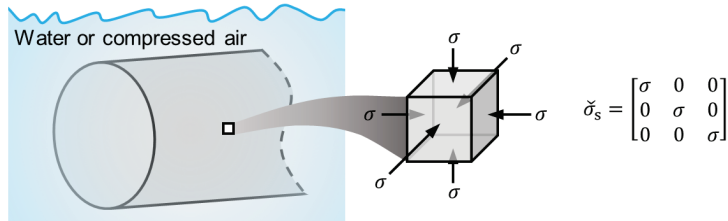

**b | Barocaloric**

**c | Twistocaloric**

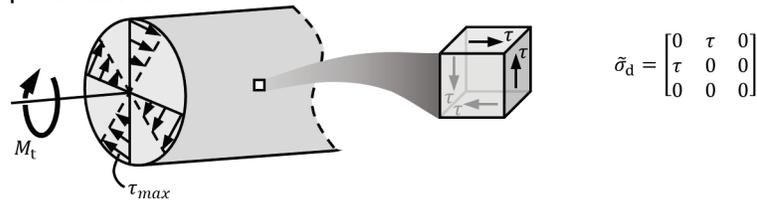

**d | Flexocaloric**

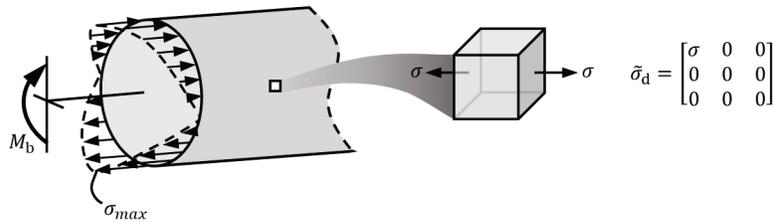

**e | Shearocaloric**

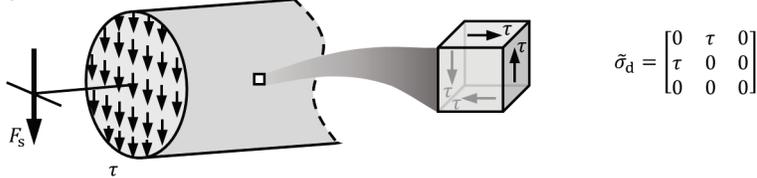

**f | Hollocaloric**

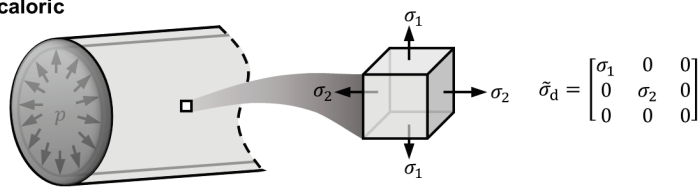

Because internal stress states of materials are often intimately coupled to other properties such as local magnetic ordering and electric dipole moments, materials that exhibit deviatocaloric effects are likely to also display other caloric properties due to multicaloric coupling. A good example is a ferroelectric material with flexoelectricity which can evince a flexocaloric effect[137].

## 5. Hysteresis versus durability in monocalorics and multicalorics

Hysteresis in the space of the field–conjugate displacement (FIG. 1i) has been widely recognized as the key parameter which governs the durability and fatigue life of caloric materials[6,11,37,138-141]. For many materials in fact the hysteresis is a direct measure of the efficacy of the first-order transition associated with the caloric effect, and now becomes a simple yet quantifiable parameter candidate in the caloric community to assess the performance of a cooling cycle and cooling devices[142-144]. In first-order phase transitions, hysteresis represents the dissipated energy in the act of changing ferroic order parameters through dissipative losses of different origins such as internal friction[145,146] or domain-wall pinning[147-150]. The conventional wisdom[148,151-153] holds that smaller hysteresis allows the phase-transforming materials to degrade less and thus can last longer. Mechanical integrity is of utmost importance to all caloric materials and their applications, and this is particularly the case for mechanocaloric materials. For electrocaloric cooling, the breakdown of electrical insulation (which may be accompanied by mechanical breakdown) is a major cause of failure and degradation. For magnetocaloric materials, aside from corrosion due to contact with the heat-exchange medium[154,155], mechanical brittleness[156] as well as fatigue and mechanical breakdown due to magnetostriction[3,140] has been discussed. Thus, mechanical properties and integrity have various roles to play in governing the degradation behavior of different types of caloric materials. As such, reducing (and ideally eliminating) hysteresis has been a central goal in the caloric materials community. We summarize representative strategies and mechanisms for hysteresis minimization in the table of Box 2. One well-accepted strategy is to tune the materials composition in order to optimize lattice compatibility and reduce mismatch strains between phases[138,157], thereby minimizing the formation of interfacial defects during transformation to extend lifetime.

Co-existing (and sometimes coupled) order parameters in multicaloric materials have unique energy landscapes with profound implications from the hysteresis point of view. In addition to intrinsic hysteresis of the materials themselves, it has been shown that multicaloric cycles can nominally suppress the extrinsic hysteresis by leveraging the sensitivity of a ferroic transformation on a non-conjugate field. For instance, the occurrence of a magnetic transition can be tuned by stress. In one example, a nominal reduction of magnetic hysteresis was observed in $Ni_{45.2}Mn_{36.7}In_{13}Co_{5.1}$ at 308 K when it was magnetized at 7 T in zero pressure and then the field was removed under a pressure of 130 MPa[9]. In another example, a $Fe_{50}Rh_{50}/BaTiO_3$ composite was first magnetized in a 5 T field with zero electric field at 385 K; subsequently, an electric field of 0.2 MV $m^{-1}$ was applied to $BaTiO_3$, and then the magnetic field was removed, followed by removal of the electric field[98], which also leads to a nominal reduction of magnetic hysteresis. We point out that in both cases the hysteresis is simply shifted to a different field variable and the attainment in minimizing the dissipative losses of energy is not clear in essence.

Box 2 | **Impact of hysteresis on functional fatigue and strategies for hysteresis reduction**

Dissipated energy percentage, $\frac{\Delta}{\Phi}$, is defined as the ratio of the dissipative losses of energy manifested as the field loop hysteresis area, $\Delta$ (blue-colored shaded region in Panel **a** and **b**) to the stored energy by input work, $\Phi$ (shaded region in Panel **a** and **b**) in first-order phase-transformation materials and measures how much the dissipative losses of energy are occupying the stored energy, which has been found to correlate well with their functional fatigue behavior. The role of the dissipative losses of energy on fatigue life has been recognized in the transformation process where reduced hysteresis can lead to extended durability (See the table within Box 2 for strategies and mechanisms). The stored energy from the work by the external fields during the forward transformations provides the potential energy for reverse transformations and facilitates the reversibility and endurance limit.

First identified in elastocaloric materials, the trend is such that minimizing the ratio, $\frac{\Delta}{\Phi}$, is key to extending functional life, $N_f$ (sustained cycles). Four materials cases are illustrated in the order of optimization level (Panel **b**): Material 1 has large $\Phi$, large $\Delta$, and large $\frac{\Delta}{\Phi}$, unfavorable for achieving a long functional life. Material 2 has medium $\Phi$, medium $\Delta$, and large $\frac{\Delta}{\Phi}$, unfavorable for long functional life. Material 3 has medium $\Phi$, small $\Delta$, and small $\frac{\Delta}{\Phi}$, favorable for long functional life. An ideal case is Material 4 with small $\Phi$, minimal $\Delta$, and minimal $\frac{\Delta}{\Phi}$, most likely to display a relatively long functional life. Low critical transformation field, $X_c$ (dashed line in Panel **a)** is highly favored for ease of driving materials to transform. From the energetics point of view, the same principle is expected to apply to other caloric processes involving magnetic, electric, or combined fields.

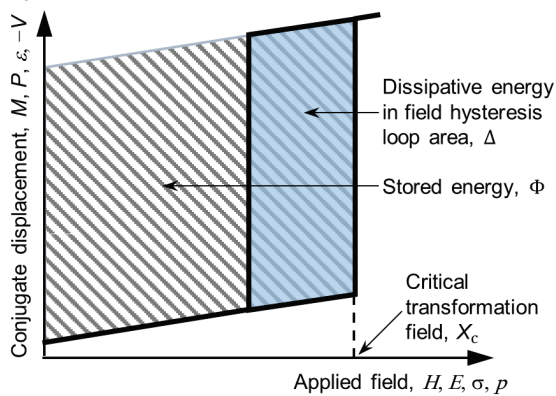

**a | Stored energy vs dissipative energy**

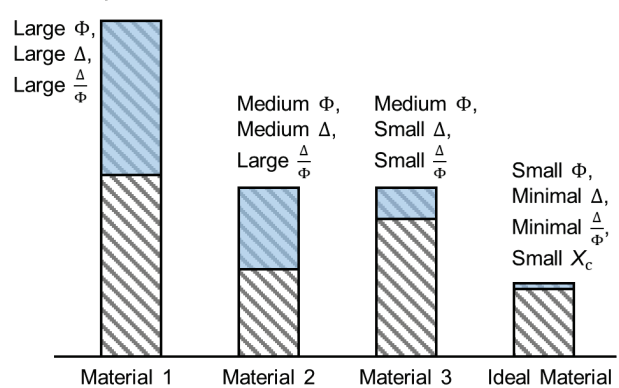

**b | Optimizing dissipated energy percentage**

| Box 2 \| **(Continued)** |||||
|---|---|---|---|
| **Hysteresis reduction measures in first-order-type caloric materials** ||||
| **Caloric process** | **Selected strategies** | **Mechanisms** | **References** |
| Elastocaloric (eC) | Tuning composition | Interface compatibility through lattice matching | 138,153,262,263 |
|  | Smoothing surfaces and purifying volumes | Reducing surface and volumetric defects to minimize pinning dissipation | 158,159,291 |
|  | Creating eutectic microstructures | Switching the transformation mode from propagation to coalescence for minimizing motional friction | 11,265,292 |
| Magnetocaloric (MC) | Alloying with new elements | Superparamagnetic behavior with suppressed crystallographic phase change | 139 |
|  | Substituting partial chemical elements | Weakening itinerant-electron metamagnetic transition | 148,293,294 |
|  | Tailoring materials geometry | Reducing volumetric strain constraints and engineering local demagnetization | 295-297 |
| Electrocaloric (EC) | Polymer blend with crosslinking | Disrupting formation of large polarization domains by miscibility-induced inter-molecular interactions | 298-300 |
|  | Unipolar cycling | Incomplete reversal for minimizing the formation of polarization domains | 141,248,301 |
| Multicaloric (MulC) | Introducing non-conjugate fields | Overlapping one original loop with another loop shifted under non-conjugate fields to form an intersected loop, although the overall dissipative losses of energy are undiminished | 9,98,99 |

The energy dissipated as hysteresis in a back-and-forth transformation process, however, is not the only mechanism that influences the functional fatigue life in phase-transforming materials. In the context of elastocaloric alloys (which are used for mechanocaloric cooling), it was recently proposed that the stored elastic energy in a forward transformation not only provides the driving force for the return transformation eliminating the need to lower the critical reversal stress[158,159], but it also plays an important role in transformation reversibility[160] (a measure of transformation efficacy) if accommodation of shape and volume changes between phases accompanies a strong slip resistance of the high-temperature phase[161], namely, a high yield strength. The endurance limit of a material subjected to stress has a monotonic correlation on mechanical strength for materials ranging from metals and polymers (with a correlation coefficient of 0.33) to ceramics and glasses (with a correlation coefficient of 0.9)[162]. Equivalent upper limits for the endurance are dielectric strength and breakdown magnetostriction for electrocaloric materials and magnetocaloric materials, respectively[3].

Dividing the dissipated energy associated with the hysteresis by the stored energy, one obtains a dimensionless quantity we call "dissipated energy percentage"[11], which provides a direct measure to monitor and predict functional life in first-order phase-transforming materials. A qualitative relation between the dissipated energy percentage, $\frac{\Delta}{\Phi}$, and functional life can be expressed as:

$$N_f \propto \left(\frac{\Delta}{\Phi}\right)^{-1} \qquad (1)$$

where $N_f$ is the sustained cycles—the number of cycles at the onset of loss of their functionality[11], $\Delta$ is the hysteresis area enclosed by the field loop, and $\Phi$ is the stored energy per unit volume per field cycle. The qualitative relation in Eq. 1 captures a trend in a series of elastocaloric materials (Supplementary Figure 1), where lowering $\frac{\Delta}{\Phi}$ was found to extend the functional life. When the $\frac{\Delta}{\Phi}$ is below a nominal threshold of ~10%, $N_f$ of more than a million can be observed[11].

The concept of dissipated energy and stored energy can be extended from elastocaloric materials to other caloric materials (Box 2) including magnetocaloric and electrocaloric materials, where research in fatigue-related properties has received increasing attention recently[140,141,163-167]. For instance, a magnetocaloric material MnFe$_{0.95}$P$_{0.595}$B$_{0.075}$Si$_{0.33}$ with a $\frac{\Delta}{\Phi}$ of 2.8% has been shown to survive more than 10,000 cycles with a $\Delta T_{ad}$ of 2.55 K[140,168]. Nano-structured polycrystalline (La$_{0.6}$Ce$_{0.4}$)$_2$Fe$_{11}$Si$_2$H$_y$ with a $\frac{\Delta}{\Phi}$ of ~1% has sustained 10$^5$ cycles with a $\Delta T_{ad}$ of 2.2 K[169], while porous polycrystalline LaFe$_{11.6}$Si$_{1.4}$ with a $\frac{\Delta}{\Phi}$ of 20.0% has been measured for 800 cycles with a $\Delta T_{ad}$ of 7.3 K[167]. An electrocaloric Pb(Mg$_{1/3}$Nb$_{2/3}$)O$_3$ relaxor with a $\frac{\Delta}{\Phi}$ of 23.0% has been cycled 70,000 times with a $\Delta T_{ad}$ of 1.2 K, while a single-crystal BaTiO$_3$ with a large $\frac{\Delta}{\Phi}$ of 83.1% starts degrading after 10 cycles[170].

The level of 10$^6$ cycles has recently been reported in a number of materials whose $\frac{\Delta}{\Phi}$ is collectively below the nominal threshold of 10%. The electrocaloric Ba(Zr$_{0.2}$Ti$_{0.8}$)O$_3$ ceramics with a $\frac{\Delta}{\Phi}$ of 9.5% (REF.[165]) via unipolar cycling of 20 kV cm$^{-1}$ exhibit a $\Delta T_{ad}$ of 0.2 K with electric-field polarity reversed every 10$^5$ cycles to maintain the $\frac{\Delta}{\Phi}$ for reaching 10$^6$ cycles. The electrocaloric PMN–10PT bulk relaxor ceramics with a $\frac{\Delta}{\Phi}$ of 5.5% (REF.[15,164]) operated with unipolar cycling of 90 kV cm$^{-1}$ can exhibit a $\Delta T_{ad}$ of 1.3 K for 10$^6$ cycles. Magnetocaloric La(Fe$_{0.865}$Co$_{0.05}$Si$_{0.085}$)$_{13}$ alloys with a $\frac{\Delta}{\Phi}$ of 7.0% (REF.[166,171]) cycled with the field of 0.6 T exhibit a $\Delta T_{ad}$ of 0.5 K for 10$^6$ cycles under a chemically protected environment against corrosion (otherwise, failed within 24 h in exposure to air or within 1 h in distilled water with 0.1 M of Na$_2$WO$_4$). Just as a benchmark, if a material used in a cooling appliance is to survive a typical life of a commercial product of ten years, it must be able to sustain performance for at least 78 million cycles operating at 1 Hz, assuming it is used 6 months per year, and 12 hours a day.

The same relation in Eq. 1 should in principle apply to second-order phase-transforming materials, where apparent minimal hysteresis translates to arbitrarily long functional life (even though *all* materials have finite operational life in practice). Reducing $\frac{\Delta}{\Phi}$ is able to extend the sustained cycles in monocaloric materials as demonstrated in Supplementary Figure 1, and in principle, can be a viable strategy for multicaloric materials to withstand stringent requirements of the million-cycle performance placed by commercial applications of cooling devices. Ideally, multicaloric materials not only reduce its nominal field hysteresis by, for example, introducing non-conjugate fields (see the table within BOX 2), but also minimize the dissipative losses of energy and the $\frac{\Delta}{\Phi}$ (panel **a** of the figure within BOX 2). It is preferable that a full transformation accompanying such reduction/minimization can be achieved by dialing back the magnitude of the applied field (panel **b** of the figure within BOX 2), which can attain the easy driving of the phase transformation in the materials and facilitate the compact design of the systems.

## 6. Technological potentials of multicalorics

Now we look at the prospect of technological advantages of multicaloric cooling with an eye toward the development of system prototypes. In Table 1, we list some of the best experimentally observed values of relevant properties from monocaloric materials, which can serve as the baseline and building blocks for multicaloric cooling techniques and systems. The potential benefits of multicaloric cooling over monocaloric cooling are the possibilities of enhancement in a multitude of properties as well as widening of operational parameters (Table 1): a boost in thermal properties (adiabatic temperature change, $\Delta T_{\text{ad}}$, isothermal entropy change, $\Delta S$, and refrigerant capacity, $RC$), widening of operating temperature window ($OTW$), nominally reduced field hysteresis ($\Delta X$), and lowered driving field.

Table 1 | **Best-reported properties of monocaloric cooling materials and systems near room temperature, and potential improvement by multicalorics (See Supplementary Table 1 for more examples of multicalorics)**

| Metrics | Best experimentally reported values for monocalorics | Multicalorics | |
|---|---|---|---|
| | | Reported examples | Potential |
| **Materials** | | | |
| Adiabatic temperature change ($|\Delta T_{\text{ad}}|$) by direct experimental measurement | 17–58 K (elastocaloric) [34,71,73,220]<br>~6.5–16.4 K (barocaloric) [77,221,222]<br>20.8 K (twistocaloric) [131]<br>8.1–15.4 K (flexocaloric) [133,134,136] | • Varying magnetic field of 0 ⇒ 1 T and triaxial stress field of 0 ⇒ 490 MPa on Ni$_{49.3}$Mn$_{34.8}$In$_{15.9}$ to attain $\Delta T_{\text{ad}}$ of 2.4 K (Category II)[108]<br>• Varying magnetic field of 0 ⇒ 10 T at constantly held uniaxial stress field of 80 MPa on Ni$_{37.0}$Co$_{13.0}$Mn$_{34.5}$Ti$_{15.5}$ to attain $\Delta T_{\text{ad}}$ of –9.0 K (Category II)[179] | ↑ |
| | 6.2–12.9 K (magnetocaloric; quasi-static fields) [223-225]<br>15.0 K by 7.5 T[226], 60.5 K by 62 T[227], 60.0 K by 55 T[228], 20.0 K by 50 T[229] (magnetocaloric; pulsed fields) | | |
| | 2.1–5.5 K (electrocaloric; bulk) [10,230-236]<br>12 K by 170 MV m$^{-1}$ (Ref.[237]), 12 K by 120 MV m$^{-1}$ (Ref.[237]), 40 K by 125 MV m$^{-1}$ (Ref.[238]), 20 K by 160 MV m$^{-1}$ (Ref.[238]) (electrocaloric; thin film) | | |
| Isothermal entropy change ($|\Delta S|$) normalized by mass | ~60–70.7 J kg$^{-1}$ K$^{-1}$ (elastocaloric) [239,240]<br>510 J kg$^{-1}$ K$^{-1}$ (barocaloric) [85] | • Varying magnetic field of 0 ⇒ 4 T at constantly held uniaxial stress field of 40 MPa on Ni$_{50}$Mn$_{35.5}$In$_{14.5}$ to attain $|\Delta S|$ of 23.9 J kg$^{-1}$ K$^{-1}$ (Category II)[107]<br>• Varying magnetic field of 0 ⇒ 5 T at constantly held electric field of –0.6 MV m$^{-1}$ on Fe–Rh/PMN–PT composite to attain $|\Delta S|$ of 15.6 J kg$^{-1}$ K$^{-1}$ (Category IV)[100] | ↑ |
| | ~18–47.3 J kg$^{-1}$ K$^{-1}$ (magnetocaloric) [31,49,52,241-246] | | |
| | ~3.1–8.0 J kg$^{-1}$ K$^{-1}$ (electrocaloric; bulk) [247-249] | | |
| | 130 J kg$^{-1}$ K$^{-1}$ (electrocaloric; thin film) [250] | | |
| Refrigerant capacity ($RC$) | ~2,300 J kg$^{-1}$ (elastocaloric) [251]<br>~2,500–2,700 J kg$^{-1}$ (barocaloric) [84,222] | Varying uniaxial stress field of 0 ⇒ 5.24 MPa at constantly held magnetic field of 0.87 T on Ni$_{43}$Mn$_{40}$Sn$_{10}$Cu$_7$ to attain $RC$ of 6.0 J kg$^{-1}$ (Category II)[110] | ↑ |
| | ~300–1,346 J kg$^{-1}$ (magnetocaloric) [139,252-254] | | |
| | 130 (electrocaloric; bulk) [255]<br>662–2,000 J kg$^{-1}$ (electrocaloric; thin film) [238,256] | | |
| Operating temperature window ($OTW$) | ~130 K (elastocaloric) [251]<br>~60 K (barocaloric) [84] | Varying magnetic field of 0 ⇒ 2 T and hydrostatic pressure of 500 ⇒ 0 MPa on Fe$_{49}$Rh$_{51}$ to attain $OTW$ of 50 K (Category II)[176] | ↑ |
| | ~40–60 K (magnetocaloric) [31,139,257-259] | | |
| | ~30–60 K (electrocaloric) [249,260,261] | | |
| Hysteresis ($\Delta X$) by first order phase transition | ~0–2 K (elastocaloric; thermal) [153,262]<br>~0–900×10$^3$ J m$^{-3}$ (elastocaloric; field loop) [11,72,263-265] | • Varying magnetic field of 0 ⇒ 7 T at constantly held hydrostatic pressure of 130 MPa on Ni$_{45.2}$Mn$_{36.7}$In$_{13}$Co$_{5.1}$ to attain the intersected field loop with the undiminished dissipative losses of energy (Category II)[9]<br>• Varying magnetic field of 0 ⇒ 5 T at constantly held electric field of 0.2 MV m$^{-1}$ on Fe–Rh/BaTiO$_3$ composite to attain the intersected field loop with the undiminished dissipative losses of energy (Category IV)[98] | ↓ |
| | ~1 K (magnetocaloric; thermal) [31]<br>~0 J m$^{-3}$ (magnetocaloric; field loop) [102,139] | | |
| | ~2 K (electrocaloric; thermal) [266]<br>~0–70×10$^3$ J m$^{-3}$ (electrocaloric; field loop) [10,141,267] | | |
| Fatigue life in number of cycles ($N_f$) | 1,000×10$^3$–100,000×10$^3$ cycles @ >4 K of $\Delta T_{ad}$ (elastocaloric) [11,263,268-270] | — | ↑ |
| | 90×10$^3$ cycles @1.9 K of $\Delta T_{ad}$ [271],<br>1,000×10$^3$ cycles @ 0.5 K of $\Delta T_{ad}$ (magnetocaloric) [166] | | |
| | 1,000×10$^3$ cycles @ ~1.5 K[141], 1.3 K[164], 0.3 K[165] of $\Delta T_{ad}$ (electrocaloric) | | |
| **Systems** | | | |
| Driving field | ~300–600 MPa (elastocaloric) [14,272-274] | Varying magnetic field of 0 ⇒ 2 T and uniaxial stress field of 0 ⇒ 80 MPa on Ni$_{49.3}$Mn$_{34.8}$In$_{15.9}$ yields a cooling effect similar to that by varying magnetic field of 0 ⇒ 4 T (Category II)[120] | ↓ |
| | ~1–1.5 T (magnetocaloric) [13,184,275-278] | | |
| | ~10–100 MV m$^{-1}$ (electrocaloric) [15,61-63,192,279] | | |
| | 65 W with 0.105 kg [76],<br>7.9 W with 0.00126 kg [280] (elastocaloric) | — | ↑ |
| Cooling power for a given mass of refrigerant | 3,042 W with 1.52 kg (magnetocaloric) [13] | | |
| | 0.64 W with 0.00023 kg [63], 0.26 W with 0.022 kg [61] (electrocaloric) | | |
| Regenerator temperature span | 15.3–19.9 K (elastocaloric) [14,281,282] | — | ↑ |
| | 33–45 K (magnetocaloric) [283-285] | | |
| | 6.6–13.0 K (electrocaloric) [61,64,192] | | |
| Operating frequency | 0.125–4 Hz (elastocaloric) [14,76,272,281,286] | — | – |
| | 0.5–4 Hz [13,182,184,275,287-289], 20 Hz [217] (magnetocaloric) | | |
| | 0.15–1.25 Hz (electrocaloric) [15,62,192] | | |

In the Materials section, listed are the metrics of those materials that exhibit the first-order phase transitions and have been experimentally measured in a direct method. In the Systems section, listed are the achieved system performance parameters among different prototypes. Within one specific caloric cooling system, the system performance depends on the operating parameters during the optimization. Due to the complex influence of the operating parameters, we qualitatively present the system coefficient of performance ($COP_{\text{system}}$) in FIG.3. Potentials for multicalorics to increase and decrease the best-reported property values are denoted by the up-arrow and down-arrow, respectively, and the dash symbol indicates no clear pathways with increase/decrease at the moment. Unless specified in other conditions, the largest values of cooling power are at zero temperature span, and the largest values of temperature span are at zero cooling power.

From Table 1, elastocaloric materials exhibit some of the largest $\Delta T_{\text{ad}}$, the widest $OTW$, and the largest $N_{\text{f}}$, and yet require the driving fields at a high level of several hundred of MPa that need to be lowered for small sizes of prototypes. They have been playing a major role in multicaloric operations. For example, using the quasi-direct method in $Ni_{50}Mn_{35.5}In_{14.5}$, a 100% increase in elastocaloric $\Delta T_{\text{ad}}$ from 1.25 K at constant 0 T to 2.5 K at constant 4 T by varying 40 MPa has been observed while the observed magnetocaloric $\Delta T_{\text{ad}}$ by varying 4 T has seen a 12.2% decrease from 4.9 K at constant 0 MPa to 4.3 K at constant 40 MPa (REF.[107]). The elastocaloric $OTW$ in PMN–32PT single crystals can be increased by 62.5% from 40 K to 65 K with an application of the electric field of 1.5 MV m$^{-1}$ (REF.[172,173]). In Ni–Mn-based alloys[174,175], one can leverage the $OTW$ of the elastocaloric effect in one temperature range and the $OTW$ of the magnetocaloric effect in another range in order to maximize their overall $OTW$. The readers are referred to Supplementary Table 1 for a complete list of multicaloric operations, which we categorize into three types: 1) the multiple external fields are varied simultaneously, 2) one external field is varied and at the same time, the other external fields are constantly held non-zero, and 3) one external field is varied and the others are zero. For single-phase and composite multicaloric materials (Table 1, Supplementary Table 1), if the $\frac{\Delta}{\Phi}$ in Eq. 1 remains below 10%, the demonstration of $\frac{\Delta}{\Phi}$ against $N_{\text{f}}$ for elastocaloric, magnetocaloric, and electrocaloric materials (Supplementary Figure 1) tells that sustained cycles of up to millions in operation can be expected.

On the other hand, the co-existence of ferroic order parameters is not always a blessing when trying to harness entropy change from both degrees of freedom through multicaloric processes. The presence of an inverse-caloric effect (due to reversed directions of entropy changes[110,176-178]) in a material can create a conflicting scenario. A case in point is an all-*d*-Heusler alloy Ni$_{37.0}$Co$_{13.0}$Mn$_{34.5}$Ti$_{15.5}$, where $\Delta T_{\text{ad}}$ of 17.7 K can be achieved with a pulsed magnetic field of 10 T in absence of stress field; with a constantly held uniaxial load of 80 MPa, however, $\Delta T_{\text{ad}}$ is now reduced to 9.0 K under the same magnetic field[179]. It is the negative sign of the cross-susceptibility between multiple degrees of freedom in such a multiferroic material that can prove to be detrimental in overall entropy change. Therefore, close inspection of the complex energy landscape governed by cross-coupled order parameters is necessary when trying to properly utilize single-phase materials (Quadrant II of FIG. 2(a)) for multicaloric cooling.

For caloric cooling technology, the road between fundamental materials science and widespread commercialization is to be paved with innovations in device engineering and prototype development. Primarily based on linear and rotary drive principles, more than 80 magnetocaloric system prototypes have been constructed worldwide to date with cooling power ranging from a few to thousands of watts[13,16,42]. In particular, magnetocaloric wine coolers were showcased jointly by Haier, Astronautics, and BASF at the International Consumer Electronics Show in Las Vegas in 2015 (REF.[180]), a comprehensive assessment of a 31-bottle wine cooler cabinet was reported in 2020 (REF.[181]), and a magnetocaloric proof-of-concept unit was piloted for a large industrial partner in 2021 (REF.[182]). There has also been a report of the operation of a 65 W elastocaloric prototype[76]. To the best of our knowledge, there is yet to be a demonstration of a system prototype based on multicaloric operation.

For a given set of operation temperatures, $T_{\text{c}}$ and $T_{\text{h}}$, which are the temperatures of the heat source and the temperature of the heat sink, respectively, one can compare the efficacy of an instrument to convert input work into cooling power, measured by the coefficient of performance (COP) for any refrigeration or heat pump system, the ratio of which to Carnot COP represents the ultimate measure of thermodynamic perfection. The state-of-the-art commercial, vapor compression systems such as factory-made air conditioners typically exhibit the system COP ($COP_{\text{system}}$) of 2–4 at $T_{\text{h}}$ = 308 K and $T_{\text{c}}$ = 300 K (air-to-air temperature in rating; AHRI Standard 210/240, United States Code of Federal Regulations (CFR) 430.32). There have been very few reports of $COP_{\text{system}}$ of caloric cooler systems[53,183,184], and it underscores the challenges associated with constructing systems and performing accurate measurements of their $COP_{\text{system}}$. The challenges of constructing systems and performing accurate measurements are made harder if multiple fields need to be applied. They include controlling the sequencing of applying multiple driving fields and its coordination with the flow of heat exchange fluid. A concept of a rotary multicaloric system has recently been proposed by considering a Halbach magnet array and a set of mechanical motors and cam tracks[185].

However, because of the latest development in caloric materials and devices as well as the advent of multicaloric cooling schemes, there are reasons to be optimistic, and here we present a systematic way (FIG. 3) to analyze and estimate the *achievable* $COP_{\text{system}}$ based on a formalism that we have previously developed[186]. All $COP$ values are to be calculated for a given set of $T_{\text{c}}$ and $T_{\text{h}}$, and they are compared against the Carnot COP ($COP_{\text{Carnot}}$), the theoretical upper limit defined by the second law of thermodynamics for delivering cooling at $T_{\text{c}}$ while rejecting heat at $T_{\text{h}}$. $COP_{\text{Carnot}}$ is 37.5 for $T_{\text{h}}$ = 308 K and $T_{\text{c}}$ = 300 K, which are temperature values used here throughout in consideration of near-room temperature applications. The first COP to consider is the materials COP ($COP_{\text{mat}}$), which is a measure of the intrinsic efficacy of the heat pumping process at the level of the materials, and it represents the starting COP which can be taken as the potential of the materials for achieving efficient processes when eventually implemented in a cooling delivery machine. Against $COP_{\text{mat}}$, we then consider various extrinsic loss mechanisms that come about due to necessary components "added" to the material. They are the field driver loss, heat transfer loss, cyclic loss, and parasitic power loss, and they all work to depreciate or erode the starting $COP_{\text{mat}}$.

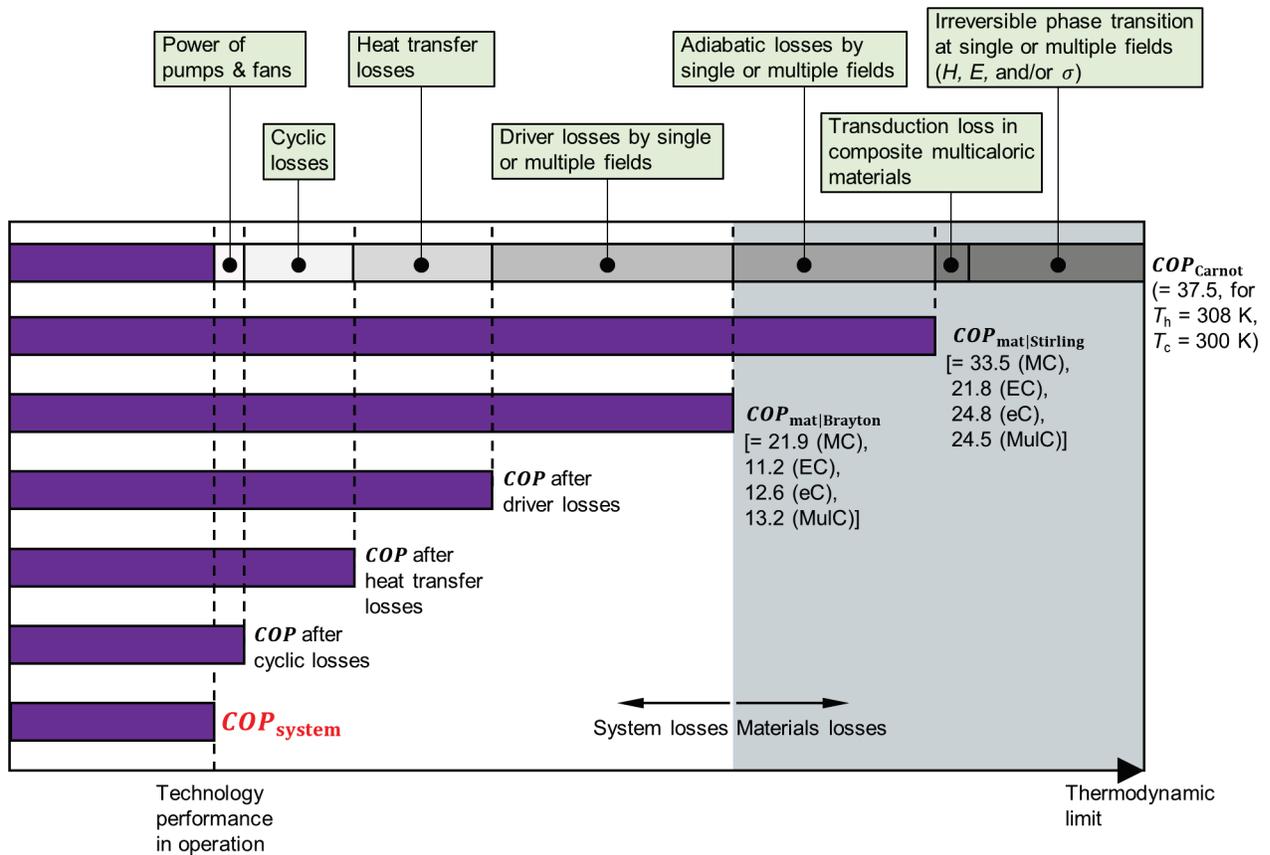

Figure 3 | **Loss factors in monocaloric and multicaloric cooling systems**. Various losses exist in functioning caloric technology. The coefficient of performance (COP) of the ideal Carnot cycle, $COP_{\text{Carnot}}$, represents the thermodynamic limit. The materials COP for the Stirling cycle, $COP_{\text{mat|Stirling}}$, considers irreversibility of field-induced phase transition and represents the efficiency potential of materials. The materials COP for the Brayton cycle, $COP_{\text{mat|Brayton}}$, takes into account the irreversibility of adiabatically applying and removing the stimulating fields to a material. The system losses comprise four components: 1) losses in drive mechanisms which generate stimuli fields, 2) heat transfer losses between caloric materials and transfer medium limit the extraction of full caloric latent heat, 3) during cyclic operations, dead thermal mass degrades transport of caloric latent heat, and 4) auxiliary parts such as pumps, fans, valves, control boards, etc, contribute to parasitic power dissipation. The final technology performance is $COP_{\text{system}}$ highlighted in red. Compared to a monocaloric system, a composite multicaloric system might have a loss due to transduction between composite components. The highest reported and estimated $COP_{\text{mat}}$ of elastocaloric (eC), magnetocaloric (MC), electrocaloric (EC), and multicaloric (MulC) materials are listed (See Supplementary Note 3 and Supplementary Table 6). The $T_h$ = 308 K and $T_c$ = 300 K are chosen based on AHRI Standard 210/240, where a typical vapor compression-based air-conditioner has a $COP_{\text{system}}$ ranging from 2 to 4.

Among the best-known thermodynamic cycles for caloric cooling, the Stirling cycle possesses a higher efficiency than other cycles, because, in the Stirling cycle, the isothermal field-variation processes and the iso-displacement regenerative processes have low irreversible losses and thus minimize the cycle area in the temperature–entropy diagram[42]. Thus, the Stirling cycle can be used to screen the caloric materials whose irreversibility in materials themselves is the loss factor compared to the Carnot cycle (FIG. 3). The $COP_{\text{mat}}$ for a material in a Stirling cycle subjected to a single external field *X* (FIG. 2d, I.1–3) can be expressed as:

$$COP_{\text{mat|Stirling},X} = \frac{T_c \Delta s_X - T_c s_{\text{gen},X}}{(T_h - T_c)\Delta s_X + (T_h + T_c)s_{\text{gen},X}} \quad (2)$$

where $\Delta s_X = \Delta s(T_0, 0 \to X)$ is the field-induced specific entropy change, $s_{\text{gen},X} = \frac{\oint_0^X dw|_{T=T_0}}{2T_0}$ is the specific entropy generation due to the irreversibility of the phase transition, $w$ is the specific work during an isothermal test at the test temperature $T_0$. In this manner, $COP_{\text{mat}}$ can be determined using basic materials properties, and experimentally, they have been shown to be as high as ≈80% of $COP_{\text{Carnot}}$ (REF.[34]). Similarly, for a multicaloric material, $COP_{\text{mat}}$ in a Stirling cycle subjected to two fields $X_1$ and $X_2$ (FIG. 2d, II.1–4) can be expressed as:

$$COP_{\text{mat|Stirling},X_1+X_2} = \frac{T_c \Delta s_{X_1+X_2} - T_c s_{\text{gen},X_1+X_2}}{(T_h - T_c)\Delta s_{X_1+X_2} + (T_h + T_c)s_{\text{gen},X_1+X_2}} \quad (3)$$

where $\Delta s_{X_1+X_2} = \Delta s(T_0, (0,0) \to (X_1, X_2))$, and $s_{\text{gen},X_1+X_2} = \frac{\oint_0^{X_1} dw_1|_{T=T_0} + \oint_0^{X_2} dw_2|_{T=T_0}}{2T_0}$, $w_1$ and $w_2$ are the specific work by field $X_1$ and field $X_2$ during an isothermal test at the test temperature $T_0$, respectively. From the viewpoint of practical operation in devices, instead of isothermal operations as shown in Eqs. 2 and 3, caloric heat pumping processes are often driven by the adiabatic application and removal of fields for ease of operation, and the Brayton cycle is most commonly applied. In addition to the losses in the materials themselves as considered in the Stirling cycle, the Brayton cycle involves cycle-related irreversibility as the additional loss that the adiabatic application and removal of fields bring in (FIG. 3). One can similarly derive $COP_{\text{mat}}$ in Brayton cycles (see Supplementary Notes 2–3 and Supplementary Tables 3–6 for derivations and data of both cycles). In FIG. 3, we list some of the largest reported values of $COP_{\text{mat}}$ for Stirling and Brayton cycles.

Because of the sheer diversity of ways in which caloric cooling are realized in device and heat exchange implementation as well as a variety of losses in systems, it is difficult to place specific numbers to the COPs after the four mechanisms of the losses which successively work to undercut $COP_{\text{mat}}$. The key message of FIG. 3, however, is that given the starting $COP_{\text{mat}}$ is as high as 20–30 for caloric materials, if we aim to end up with a $COP_{\text{system}} > 4$, the difference represents challenges and opportunities for the development of highly efficient systems that can potentially rival vapor-compression systems.

### 7. Future perspectives

Two key factors which continue to drive and motivate innovations in cooling technologies are high efficiency and minimization of global warming potentials in refrigerants. The advances in multicaloric materials and devices can potentially enable solid-state cooling technology to cross over the technological threshold for thriving commercialization. Just as the development in magnetocaloric materials in the late 1990s fueled their widespread scientific pursuit and industrial interests[49,187,188], we believe multicaloric cooling today stands at the cusp, poised to lead the next decades of innovations in caloric cooling at large.

We envisage the seeds of advances in multicaloric cooling will be sown in the forms of improved materials properties such as reduction of hysteresis as well as optimization of application schemes for multiple external fields. As the concept of active regeneration first developed for magnetocaloric cooling[48,189,190] has now been adopted by the other caloric cooling techniques[14,61,191,192], we expect active multicaloric regeneration and even more advanced cross-disciplinary approaches to serve as the major launching pad of novel cooling system prototypes. From additive manufacturing to the intelligent design of heat exchangers, advanced materials processing techniques are expected to be the lynchpin for active solid refrigerants as well as components for system assembly.

Caloric and multicaloric cooling systems are expected to become more popular and viable with the potentials to penetrate the commercial market because of the existing and impending global regulations of hydrofluorocarbons and limited options for suitable replacements[193,194]. In particular, when the operational footprint of caloric cooling becomes scalable, it is expected to be competitive in small-sized applications, where a volume-specific latent heat from phase transition in caloric cooling is extremely favorable compared to vapor compression. For instance, the latent heat of a $Ni_{49.8}Ti_{30.2}Hf_{20}$ alloy[74,195] is hundreds of J cm$^{-3}$, which is orders of magnitude higher than that of vapor compression (single-digit J cm$^{-3}$ in R-134a and ~10 J cm$^{-3}$ in R-32, the typical commercial refrigerants, normalized by the specific volume of the vapor phase, REF.[196]). A high value, or more technically, a high volumetric refrigeration capacity[193], indicates the potentials for the compactness level of the system.

Beyond the mainstream Heating, Ventilation, and Air Conditioning (HVAC) applications, multicaloric cooling may find transformative applications in emerging high-tech fields. One example is the implementation of a magneto-elastocaloric effect (FIG. 2d, III.1) for remote focal brain cooling for the treatment of epileptic seizures[101] (Hou, Takeuchi, et al., non-provisional U.S. patent application publication, 2020/0096240A1) aligned with the potentials of Brain-Machine Interfaces (BMI) advanced by NeuraLink[197]. The footprint of the demonstrated device is approximately 11.5 × 2.8 × 2.8 cm$^3$ and is smaller than a human brain. The compactness of the magneto-elastocaloric device is enabled by the ultra-low magnitude of the magnetic field, ~0.16 T, where one is able to do away with the expensive and cumbersome magnets that are typically required for the high magnetic field (typically >1.5 T) of magnetocaloric materials. The key which enables the relatively compact dimensions is the fact that in this envisioned application of focal brain cooling, only a small spot in the neocortex needs to be cooled with a magnitude of ≈10 K for a short period of time[198-200]. It is the compact footprint of such a multicaloric device configuration that enables entirely new applications of cooling. Another possible application of the magneto-elastocaloric effect is hydrogen liquefaction, where the presently pursued active regenerative liquefier based on a pure magnetocaloric effect using high magnetic fields (1.8 T[201], 3 T[202], 3.3 T[203,204], 4 T[205], 6 T[206]) might be realized by lower fields for cost-effective and efficient liquefaction using the composite device. The mechanical field has been shown to contribute to the reduction of a high magnetic field by partnering with a low magnetic field to achieve a similar caloric effect[120].

**Additive manufacturing of caloric materials and devices**
Additive manufacturing (AM), where a digital design guides a stream of an energy source to consolidate feedstock materials layer-by-layer to build a three-dimensional object[207-209], can transform the synthesis and processing of caloric materials. AM can overcome the challenges associated with complex regenerator architectures with a large surface-to-volume ratio desirable for efficient heat transfer. It should also help accommodate demanding geometrical constraints imposed by the need for applications of multiple fields in composite multicaloric processes (FIG. 2d., IV.1–4).

There have already been examples where AM has seen implementation in advancing caloric technology. Selective laser melting has been used to print out the wavy-channel blocks and fin-shaped rods of magnetocaloric $La(Fe,Co,Si)_{13}$ (REF.[166]) and the double corrugated flow structures and straight flow channels of magnetocaloric $La_{0.84}Ce_{0.16}Fe_{11.5}Mn_{1.5}Si_{1.3}H_x$ (REF.[210]). Directed energy deposition technique offers the advantages in stoichiometric control of feedstock, and has been used to successfully engineer the eutectic nanocomposite microstructure of Ni–Ti-based elastocaloric materials to achieve high materials efficiency as well as extended fatigue life[11]. Directed energy deposition has recently been used to make $AlFe_2B_2$ compounds with magnetocaloric effect as well as rod and honeycomb structures[211]. We believe that additive manufacturing can not only surmount the challenge of caloric materials geometries due to limited options in wires, plates, spheres, and cylinders from traditional fabrication methods, but may also offer potential solutions to some long-standing issues in materials science such as brittleness in magnetocaloric alloys by enabling precise manipulation of microstructure using a rapid cooling rate in a highly-localized region. In such a manner, advances in AM will be central to the design, development, and deployment of essential components of multicaloric cooling systems.

We also envision that innovation will be made possible by integrating AM with artificial intelligence[212-214] to turn it into an autonomous fabrication framework, where the objective of topological optimization in multicaloric regenerators will be addressed by coordination of machine learning algorithms and robotic hardware with the capacity to *in situ* monitor and qualify its performance with an iteratively-developed database. Such a system can autonomously provide feedback with physics- and data-based models in a closed-loop manner.

**Regenerative multicaloric cooling processes**

Active regeneration bridges the limited materials $\Delta T_{\text{ad}}$ with required temperature span of practical cooling devices[48], preferably 30 K or above (fluid-to-fluid temperature across the caloric regenerators), by periodically storing and transferring heat between a heat-transfer fluid and a regenerator made of caloric materials. Active regeneration establishes a length-wise temperature profile by coupling the internal thermodynamic cycles of caloric materials arranged in series[42] and offers hitherto the best solution to building a multi-fold enhancement in temperature span in cooling systems. Advanced multicalorics with measured $\Delta T_{\text{ad}}$ of more than 30 K in a single material (e.g. Ni–Mn–Ti alloys, see Table 1) has the potential to completely revolutionize solid-state cooling with untold cooling capabilities through their implementation in active regenerators. However, such a large starting $\Delta T_{\text{ad}}$ would require new mechanisms of pumping heat-transfer fluids for the heat transfer. In active caloric regenerators, the working mechanism of pumping heat-transfer fluids for the convective heat transfer fundamentally limits the operation frequency to several Hz due to two main factors[41,215]: 1) the high pumping-work input as a consequence of the high velocity of the oscillation and the viscous fluid, and 2) the need for convective heat transfer between the heat-transfer fluid and the caloric material. Because the power output of a regenerator is proportional to the operating frequency, without advances in heat-transfer processes, the ultimate power output would be compromised[42,181], and the capabilities of the large $\Delta T_{\text{ad}}$ in materials cannot be fully harnessed.

One line of development of heat transfers in caloric devices might be thermal control elements, for example, in the form of the thermal switch[215,216]. The control of the direction and possibly the intensity of heat flux in a short interval, ~millisecond, could significantly enhance the operating frequency and at the same time maintain the efficiency achieved in the active regeneration. The activation of a thermal switch by an external magnetic, electric, mechanical field or their combination could, in principle, be designed in concert with the triggering of multicaloric effects to realize an anisotropic heat flow deployed in the single-stage, cascading, or active regenerative schemes.

Another line of development is latent heat transfer by condensation and evaporation, where the transfer rate between heat exchangers and caloric materials could become an order of magnitude faster than that of conduction or convection[217,218] and thus can enable high operating frequency comparable to the vapor compression. There have been magnetocaloric prototype demonstrations with a 10-fold improvement in cycle frequency and specific cooling power by the implementation of evaporation and condensation of degassed methanol[217,219]. Such an approach indicates that promising advanced latent heat transfer processes might also be available for various multicaloric cooling techniques.

**Outlook**

In less than 10 years after its inception, multicaloric cooling has become an exciting new direction of research in solid-state cooling with potentials for high energy-conversion efficiencies in addition to providing an entirely green platform. Multicaloric cooling derives its functionalities from the rich materials physics of multiferroic materials, and its diverse range of embodiments discussed here represent possibilities for much-needed compact earth-friendly refrigeration technologies as well as the development of completely new thermal device applications beyond conventional cooling technologies. Along with this nascent technology, there are some areas that innovations can be expected to lead to lowered costs.

**Acknowledgments**
We acknowledge useful discussions with Jun Cui, Reinhard Radermacher, Yunho Hwang, Jan Muehlbauer, David Catalini, Dongsheng Wen, Rui Bao, Yufeng Xing, Weixing Yuan, and Lifen Yuan. H.H. was supported by the National Natural Science Foundation of China (NSFC Grant No. 12002013) and the Fundamental Research Funds for the Central Universities (Grant No. 501LKQB2020105028). I.T. was supported by the U.S. Department of Energy under DE-EE0009159. S.Q. was supported by the National Natural Science Foundation of China (NSFC Grant No. 51976149), the Young Elite Scientists Sponsorship Program of CAST (Grant No. 2019QNRC001), and the China Postdoctoral Science Foundation (CPSF Grant No. 2020M683471).

**Author contributions**
H.H., S.Q., and I.T. developed the concepts and framework of this Perspective and translated them into display items and main text.

**Competing interests**
I.T. is a founder of Maryland Energy & Sensor Technologies, a company that works on elastocaloric technologies.

**Related links**
**Refrigeration and air-conditioning – Consumers:**
<http://www.environment.gov.au/protection/ozone/rac/consumers>